\pdfoutput=1
\documentclass{ws-rv975x65}
\usepackage{anonchap}
\usepackage{subfigure}     
\usepackage[square]{ws-rv-van}     
\makeindex
\renewcommand*\thesection{\arabic{section}}
\renewcommand*\thefigure{\arabic{figure}}

\begin{document}

\chapter*{The Discovery of the Gluon}

\author[John Ellis]{John Ellis}

\address{Theoretical Particle Physics and Cosmology Group, Department of
  Physics, King's~College~London, London WC2R 2LS, United Kingdom\\
\& Theory Division, CERN, CH-1211 Geneva 23,
  Switzerland, \\
John.Ellis@cern.ch}

\begin{abstract}
Soon after the postulation of quarks, it was suggested that they interact via
gluons, but direct experimental evidence was lacking for over a decade.
In 1976, Mary Gaillard, Graham Ross and the author suggested searching for the
gluon via 3-jet events due to gluon bremsstrahlung in $e^+ e^-$ collisions. Following
our suggestion, the gluon was discovered at DESY in 1979 by TASSO and the other experiments
at the PETRA collider. \\ \\
\emph{Contribution to the book `50 Years of Quarks' to be published by World Scientific}\\ \\
{\tt KCL-PH-TH/2014-36, LCTS/2014-35, CERN-PH-TH/2014-178}

\end{abstract}
\body

\section{Quarks are Not Enough}\label{ra_sec1}

When Murray Gell-Mann postulated quarks~\cite{MGM} and George Zweig postulated aces~\cite{GZ} in 1964 as the fundamental
constituents of strongly-interacting particles, several questions immediately came to mind.
What are the super-strong forces that bind them inside baryons and mesons?
Are these forces carried by other particles, perhaps analogous the photons that
bind electrons to nuclei inside atoms? But if so, why do we not see individual quarks?
What is the important intrinsic difference between the underlying theory
of the strong interactions and quantum electrodynamics (QED), the unified quantum theory of
electricity and magnetism?

It was natural to suppose that the forces between quarks were mediated by photon-like
bosons, which came to be known as gluons. If so, did they have spin one, like the photon?
If so, did they couple to some charge analogous to electric charge? It was suggested that
quarks carried a new quantum number called colour~\cite{colour}, which could explain certain puzzles
such as the apparent symmetry of the wave functions of the lightest baryons, the rate for
$\pi^0 \to 2 \gamma$ decay and (later) the rate for $e^+ e^- \to$ hadrons. However,
the postulation of colour raised additional questions. For example, did the quarks of
different colours all have the same electric charge, and could individual coloured
particles ever be observed, or would they remain confined inside hadrons?

Many of these issues were still being hotly debated when I started research in 1968.
My supervisor was Bruno Renner, who had recently co-authored with Gell-Mann and
Bob Oakes an influential paper~\cite{GOR} on the pattern of breaking of the approximate chiral
SU(3)$\times$SU(3), flavour SU(3) and chiral SU(2)$\times$SU(2) symmetries of hadrons.
I learnt from him that the gluons were presumably vector particles like photons,
because otherwise it would be difficult to understand the successes of approximate
chiral symmetry, but this posed another dilemma. On the one hand, if they were Abelian vector gluons as in QED,
how could they confine quarks? On the other hand, non-Abelian gauge theories were known,
but nobody knew how to calculate reliably their dynamics.

It was in 1968 that the results from deep-inelastic electron-proton scattering experiments at SLAC
first became known~\cite{SLAC}. Their (near-)scaling behaviour was 
interpreted by James Bjorken~\cite{Bj} and Richard Feynman~\cite{RF}
in terms of point-like constituents within the proton, called partons. It was natural to suppose
that the partons probed by electrons though virtual photon exchange
might be quarks, and this expectation was soon supported by
measurements of the ratio of longitudinal and transverse scattering~\cite{CG}.
However, this insight immediately raised other questions. Might the proton contain
other partons, such as gluons, that were not probed directly by photons? And what
was the origin of the bizarre dynamics that enabled quarks to resemble quasi-free
point-like particles when probed at short distances in the SLAC experiments, yet
confined them within hadrons?

Chris Llewellyn-Smith pointed out that one could measure the total fraction of the
proton momentum carried by quark partons~\cite{CHLS}, and the experiments showed
that this fraction was about half the total. Either the parton model was crazy, or
the remaining half of the proton momentum had to be carried by partons without
electric charges, such as neutral gluons. This was the first circumstantial
evidence for the existence of gluons.

\section{The Theory of the Strong Interactions}

In 1971, Gerardus 't Hooft gave convincing arguments that unbroken (massless)
non-Abelian gauge theories were renormalizable~\cite{tH}. In retrospect, this should have
triggered extensive theoretical studies of a non-Abelian gluon theory based on the
SU(3) colour group~\cite{BFM}, but attention was probably distracted initially by 't Hooft's extension
(together with Martinus Veltman) of his proof to spontaneously-broken (massive)
non-Abelian gauge theories~\cite{tHV}, which opened the way to renormalizable theories
of the weak interactions~\cite{SWEW}.

Giorgio Parisi realized that the key to constructing a field theory of the strong
interactions would be asymptotic freedom~\cite{GP}, the property of renormalization
driving the coupling of a field theory to smaller (larger) values at shorter (larger)
distances, but the theories studied at that point did not have this property.
Kurt Symanzik gave a talk at a conference in Marseille in the summer of 1972
where he pointed out that $\phi^4$ theory would be asymptotically free if
its coupling had an unphysical negative value~\cite{KS}. After his talk, 't Hooft remarked
that the coupling of a non-Abelian gauge theory would be driven towards zero,
but apparently neither he, Symanzik nor anybody else who heard the remark
made the connection between gauge theory and the strong interactions.

This connection was made in 1973 by David Politzer~\cite{DP} and by David Gross and
Frank Wilczek~\cite{GW}, who not only demonstrated the asymptotic freedom of
non-Abelian gauge theories but also proposed that the strong interactions be
described by non-Abelian gluons whose interactions were specified by an
unbroken SU(3) colour group. This proposal was supported strongly by
Harald Fritzsch, Gell-Mann and Heinrich Leutwyler in a paper published
later in 1973~\cite{FGL}, and came to be known as Quantum Chromodynamics (QCD).

It was possible to calculate in QCD logarithmic deviations from scaling in
deep-inelastic scattering~\cite{GPGW}. It was argued in 1974 that early results from
deep-inelastic muon scattering at Fermilab were qualitatively consistent
with these calculations~\cite{London}, since they suggested the characteristic  features
of falling structure functions at large quark-parton momentum fractions
and rising structure functions at low momentum fractions.

However, the picture was clouded by the unexpected behaviour of the
total cross section for $e^+ e^- \to$ hadrons, which was much larger
at centre-of-mass energies above 4~GeV than was calculated in QCD
with the three known flavours of quark~\cite{JELondon}. The answer was provided in late
1974 by the discovery of the charm quark~\cite{RT}, which boosted the
$e^+ e^- \to$ hadrons cross section, as observed.

\section{Where are the Gluons?}

In the mid-1970s QCD was widely regarded as the only candidate theory of the strong interactions.
It was the only realistic asymptotically-free field theory, capable of
accommodating the (approximate) scaling seen in deep-inelastic scattering. QCD had
qualitative success in fitting the emerging pattern of scaling violations through quark-parton
energy loss via gluon bremsstrahlung and gluon splitting into quark-antiquark pairs. Moreover, QCD could
explain semi-quantitatively the emerging spectrum of charmonia and had successes in calculating their decays
in terms of quark-antiquark annihilations into two or three gluons~\cite{AP}.
No theorist seriously doubted the existence of the gluon, but direct proof of its existence,
a `smoking gluon', remained elusive.

In parallel, jet physics was an emerging topic. Statistical evidence was found for two-jet events in low-energy
electronÐpositron annihilation into hadrons at SPEAR at SLAC, which could be interpreted as $e^+ e^- \to {\bar q} q$~\cite{Hanson}.
It was expected that partons should manifest themselves in high-energy hadron-hadron collisions via the
production of pairs of large transverse-momentum jets. However, these were proving elusive at the
Intersecting Storage Rings (ISR), CERN's pioneering proton-proton collider.
It was known that the transverse-momentum spectrum of individual hadron production had a tail above the
exponential fall-off seen in earlier experiments, but the shape of the spectrum did not agree with the
power-law fall-off predicted on the basis of the hard scattering of quarks and gluons,
and no jets were visible. Thus, there were no signs of quarks at the ISR, let alone gluons,
and rival theories Ð such as the constituent-interchange model~\cite{CIM}, in which the scattering objects were
thought to be mesons Ð were being advocated.

\section{The Three-Jet Idea}

\begin{figure}[htb]
\centerline{\includegraphics[width=8cm]{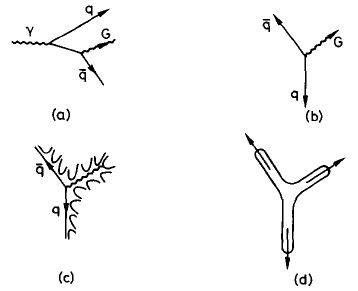}}
\caption{\it (a) A Feynman diagram for gluon bremsstrahlung. (b) Wide-angle gluon bremsstrahlung
visualized in momentum space in the centre of mass. (c) Hadronization of wide-angle gluon
bremsstrahlung. (d) The resulting 3-jet final state. Figure from Ref.~[25].} \label{Mercedes}
\end{figure}

This was the context in 1976 when I was walking over the bridge from the CERN cafeteria
back to my office one day. Turning the corner by the library, the thought occurred that the 
simplest experimental situation to search directly for the gluon would be through production via 
bremsstrahlung in electronÐpositron annihilation: $e^+ e^- \to {\bar q} q g$. What could be simpler?
Together with Graham Ross and Mary Gaillard, we calculated the gluon bremsstrahlung process in QCD~\cite{EGR}, 
see Fig.~\ref{Mercedes}, arguing that under the appropriate kinematic conditions at high energies, asymptotic freedom
would guarantee that this would be the leading correction to the process $e^+ e^- \to {\bar q}q$ that
gave two-jet final states. We showed that gluon bremsstrahlung would first manifest itself via broadening
of one of the jets in a planar configuration, and that the appearance of three-jet events would
provide the long-sought `smoking gluon', as seen in part (d) of Fig.~\ref{Mercedes}.
We also contrasted the predictions of QCD with a `straw-man'
theory based on scalar gluons.

Two higher-energy collider projects were in preparation at the time,
PETRA at DESY and PEP at SLAC, and we estimated that they should have sufficient energy to
observe clear-cut three-jet events. I was already in contact with experimentalists at DESY, 
particularly my friend the late Bj{\o}rn Wiik,
a leading member of the TASSO collaboration who was infected by our enthusiasm for the three-jet idea.
Soon after Mary, Graham and I had published our paper, I made a trip to DESY to give a seminar about it.
The reception from the DESY theorists of that time was one of scepticism, bordering on hostility,
and I faced fierce questioning why the short-distance structure of QCD should survive the hadronization process.
My reply was that hadronization was expected to be a soft process involving small exchanges of momenta,
as seen in part (c) of Fig.~\ref{Mercedes},
and that the appearance of two-jet events at SPEAR supported the idea.
At the suggestion of Bj{\o}rn Wiik, I also went to the office of G{\" u}nter Wolf,
another leader of the TASSO collaboration, to discuss with him the three-jet idea.
He listened much more politely than the theorists.

The second paper on three-jet events was published in mid-1977 by Tom Degrand, Jack Ng and Henry Tye~\cite{DNT},
who contrasted the QCD prediction with that of the constituent-interchange model and a quark-confining
string model. Then, later in 1977, George Sterman and Steve Weinberg published an influential paper~\cite{SW}
showing how jet cross-sections could be defined rigorously in QCD with a careful treatment of infrared and 
collinear singularities. In our 1976 paper we had contented ourselves with showing that these were 
unimportant in the three-jet kinematic region of interest to us. The paper by Sterman and Weinberg
opened the way to a systematic study of variables describing jet broadening and multi-jet events, 
which generated an avalanche of subsequent theoretical papers~\cite{SphT}. In particular, Alvaro De R{\'u}jula, 
Emmanuel Floratos, Mary Gaillard and I wrote a paper in early 1978~\cite{DEFG} showing how `antenna patterns' of gluon radiation 
could be calculated in QCD and used to extract statistical evidence for gluon radiation, 
even if individual three-jet events could not be distinguished.

Meanwhile, the PETRA collider was being readied for high-energy data-taking with its four detectors, 
TASSO, JADE, PLUTO and Mark J. Bj{\o}rn Wiik, one of the leaders of the TASSO collaboration, and I
were in regular contact. He came frequently to CERN around that time for various meetings,
and I was working with him on electron-proton colliders. He told me that Sau Lan Wu had joined the
TASSO experiment, and asked my opinion (which was enthusiastic) on the suggestion
that she prepare a three-jet analysis for the collaboration. In early 1979 she and Haimo Zobernig
wrote a paper~\cite{WZ} describing an algorithm for distinguishing three-jet events.

Another opportunity for detecting gluon effects had appeared in 1977, with the discovery of the $\Upsilon$
bottomonium states~\cite{Ups}. According to perturbative QCD, the dominant decay of the vector $\Upsilon(1S)$ state
should be into three gluons, so the final state should look different from the dominant two-jet final states
seen in the $e^+ e^- \to {\bar q}q$ continuum~\cite{Ups3}. A difference was indeed seen~\cite{PLUTO}, providing important
circumstantial evidence for the gluon.

\section{Proof at Last}

During the second half of 1978 and the first half of 1979, the machine team at DESY was
increasing systematically the collision energy of PETRA, getting into the energy range where
we expected three-jet events due to gluon bremsstrahlung to be detectable.
The first three-jet news came in June 1979 at the time of a neutrino conference in Bergen, Norway. 
The weekend before that meeting I was staying with Bj{\o}rn Wiik at his father's house beside a fjord
outside Bergen, when Sau Lan Wu arrived over the hills bearing printouts of the first three-jet event. 
Bj{\o}rn included the event in his talk at the conference~\cite{BW} and I also mentioned it in mine~\cite{JEB}. 
I remember Don Perkins asking whether one event was enough to prove the existence of the gluon:
my tongue-in-cheek response was that it was difficult to believe in eight gluons on the strength of a single event!
More scientifically, a single event might always be a statistical fluctuation, and more events would
be needed to cement the interpretation.

\begin{figure}[htb]
\centerline{\includegraphics[width=8cm]{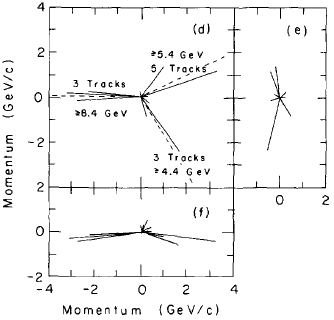}}
\caption{\it An event detected by the TASSO collaboration~\protect\cite{BW,TASSO},
exhibiting the planar 3-jet structure predicted in~\protect\cite{EGR}. Figure from Ref.~[36].} \label{3jet}
\end{figure}

The next outing for three-jet events was at the European Physical Society conference in Geneva in July 1979.
Three members of the TASSO collaboration, Roger Cashmore, Paul S{\"o}ding and G{\" u}nter Wolf,
spoke at the meeting and presented several clear three-jet events~\cite{TASSO}. The hunt for gluons was looking good!

The public announcement of the gluon discovery came at the Lepton/Photon Symposium held at Fermilab in August 1979.
All four PETRA experiments showed evidence: JADE and PLUTO followed TASSO in presenting evidence for jet 
broadening and three-jet events as suggested in our 1976 paper, while the Mark J collaboration led by Sam Ting
presented an analysis of antenna patterns along the lines of our 1978 paper.
There was a press conference at which one of the three-jet events was presented,
and a journalist asked which jet was the gluon. He was told that the smart money was on the jet on the left (or was it the right?). Refereed publications by TASSO~\cite{TASSO} and the other PETRA collaborations~\cite{PETRA} soon appeared,
and the gluon finally joined the Pantheon of established particles as the second gauge boson to be discovered,
joining the photon.

\section{After the Discovery}

An important question remained: was the gluon a vector particle, as predicted by QCD and everybody expected,
or was it a scalar boson as nobody expected? Back in 1978, Inga Karliner and I had written
a paper~\cite{EK} that proposed a method for distinguishing the two possibilities by looking at a distribution
in the angles between the three jets, based on our intuition about the nature of gluon bremsstrahlung.
This method was used in 1980 by the TASSO collaboration to prove that the gluon was indeed a vector particle~\cite{spin},
a result that was confirmed by the other experiments at PETRA in various ways.

Studies of gluon jets have developed into a precision technique for testing QCD.
One-loop corrections to three-jet cross-sections were calculated by Keith Ellis, Douglas Ross and Tony Terrano in 1980~\cite{ERT}.
They have subsequently been used, particularly by the LEP collaborations working at the $e^+ e^- \to Z^0$ peak,
to measure the strong coupling. The LEP collaborations also used four-jet events to verify the QCD predictions
for the three-gluon coupling, a crucial consequence of the non-Abelian nature of QCD.
More recently, two-loop corrections to three-jet production in $e^+ e^-$ annihilation
have been calculated~\cite{Glover}, enabling jet production to become a high-precision tool for
testing QCD predictions and measuring accurately the QCD coupling and its
asymptotically-free decrease with energy~\cite{PDG}.

Large transverse-momentum jets in hadron-hadron collisions emerged clearly at the CERN
proton-antiproton collider in 1982~\cite{UA2}. Detailed studies of gluon collisions and jets then went
on to become a staple of tests of the Standard Model at subsequent colliders. Most recently,
the dominant mode of production of the Higgs boson at the LHC is via gluon-gluon
collisions~\cite{H}.

The discovery of the gluon is a textbook example of how theorists and experimentalists,
working together, can advance knowledge. The discovery of the Higgs boson at the LHC
has been another example for the textbooks, with the gluon progressing from a discovery
to a tool for making further discoveries. 

\section*{Acknowledgments}
The author thanks Chris Llewellyn-Smith for his comments on the text.
This work was supported in part by the London Centre for Terauniverse Studies
(LCTS), using funding from the European Research Council via the Advanced Investigator
Grant 267352 and from the UK STFC via the research grant ST/J002798/1.

\end{document}